\def\Title{Error-disturbance relations in mixed states}
\def\Author{Masanao Ozawa}
\def\Affiliation{Graduate School of Information Science,
Nagoya University, Chikusa-ku, Nagoya, 464-8601, Japan}
\documentclass[prl,reprint,twocolumn,showpacs]{revtex4}
\usepackage{graphicx,bm,float}
\usepackage{amsmath,amssymb,amsthm} 
\usepackage{times}
\newcommand{\beq}{\begin{equation}}
\newcommand{\eeq}{\end{equation}}
  \newcommand{\beql}[1]{\begin{equation}\label{eq:#1}}
  \newcommand{\beqa}{\begin{eqnarray}}
  \newcommand{\eeqa}{\end{eqnarray}}
  \newcommand{\beqas}{\begin{eqnarray*}}
  \newcommand{\eeqas}{\end{eqnarray*}}
  \newcommand*{\bC}{\mathbf{C}}
  \newcommand*{\bP}{\mathbf{P}}
  \newcommand*{\bS}{\mathbf{S}}
  \newcommand*{\cA}{\mathcal{A}}
  \newcommand*{\cB}{\mathcal{B}}
  \newcommand*{\cE}{\mathcal{E}}
  \newcommand*{\cH}{\mathcal{H}}
  \newcommand*{\cK}{\mathcal{K}}
 \newcommand*{\cL}{\mathcal{L}}
  \newcommand*{\cW}{\mathcal{W}}
  \newcommand*{\al}{\alpha}
  \newcommand*{\be}{\beta} 
  \newcommand*{\da}{\dagger}
 \newcommand*{\de}{\delta}
  \newcommand*{\nn}{\nonumber}
  \newcommand*{\ph}{\phi}
 \newcommand*{\ps}{\psi} 
  \newcommand*{\rh}{\rho}
  \newcommand*{\De}{\Delta}                                          
  \newcommand*{\Eq}[1]{Eq.~(\ref{eq:#1})}                                     
  \newcommand*{\Ps}{\Psi}                                                                                  
  \newcommand*{\Tr}{\mbox{\rm Tr}}
  \newcommand*{\eq}[1]{(\ref{eq:#1})}
\newcommand*{\bra}[1]{\langle#1|}
\newcommand*{\ket}[1]{|#1\rangle}
\newcommand*{\braket}[1]{\langle#1\rangle}
\newcommand*{\bracket}[1]{\langle#1\rangle}
\newcommand*{\ketbra}[1]{\ket{#1}\bra{#1}}
\newtheorem{Theorem}{Theorem}
\newenvironment{Proof}{\begin{trivlist}
 \item[\hskip \labelsep {\em \indent Proof.}]}{\qed\end{trivlist}}
\newcommand{\ep}{\varepsilon}
\newcommand{\et}{\eta}
\newcommand{\hep}{\hat{\varepsilon}}
\newcommand{\het}{\hat{\eta}}
\newcommand{\si}{\sigma}
\renewcommand{\Re}{{\rm Re}}
\renewcommand{\Im}{{\rm Im}}

\newcommand{\kxi}{\ket{\xi}}
\newcommand{\kps}{\ket{\psi}}

\newcommand{\kb}{\ketbra}
\newcommand{\sq}{\sqrt{\rh}}
\newcommand{\ip}[2]{(#1,#2)}
\newcommand{\n}[1]{\|#1\|}
\renewcommand{\a}{\mathbf{a}}
\renewcommand{\b}{\mathbf{b}}
\newcommand{\x}{\mathbf{m}}
\newcommand{\y}{\mathbf{n}}
\newcommand{\xx}{\bar{\mathbf{m}}}
\newcommand{\yy}{\bar{\mathbf{n}}}
\newcommand{\X}{\cA}
\newcommand{\Y}{\cB}
\newcommand{\keta}{\ket{\eta}}
\newcommand{\kbxi}{\ketbra{\xi}}

\begin{document}

\title{\Title}
\author{\Author}
\affiliation{\Affiliation}
\begin{abstract}
Heisenberg's uncertainty principle was originally formulated in 1927 as a quantitative 
relation between the ``mean error'' of a measurement of one observable and
the disturbance thereby caused on another observable.
Heisenberg derived this famous relation under an additional assumption on quantum
measurements that has been abandoned in the modern theory,
and its universal validity was questioned in a debate on the sensitivity limit 
to gravitational-wave detectors in 1980s.
A universally valid form of the error-disturbance relation was shown to be derived 
in the modern framework of general quantum measurements in 2003.
We have experienced a considerable progress in theoretical and experimental study
of error-disturbance relations in the last decade.
In 2013 Branciard showed a new stronger form of universally valid error-disturbance 
relations, one of which is proved tight for spin measurements carried out in ``pure'' states.  
Nevertheless, a recent information-theoretical study of error-disturbance relations 
has suggested that Branciard relations can be considerably strengthened for measurements in mixed states.
Here, we show a method for strengthening Branciard relations in mixed
states and derive several new universally valid and stronger error-disturbance relations 
in mixed states.  In particular, it is proved that one of them gives an ultimate 
error-disturbance relation for spin measurements, which is tight in any state.
The new relations will play an important role in applications to 
state estimation problems including quantum cryptographic scenarios.
\end{abstract}
\maketitle

{\em Heisenberg's error-disturbance relation.}---The discovery of quantum mechanics 
introduced non-commutativity in physical quantities: 
the commutation relation
\beql{CCR}
[Q,P]=i\hbar
\eeq
holds between a coordinate $Q$ of a particle and its momentum $P$
\footnote{The commutator $[Q,P]$ is defined by  $[Q,P]=QP-PQ$.}.
In 1927, Heisenberg found 
an operational meaning of the non-commutativity:
``the more precisely the 
position is determined,  the less precisely the momentum is known, and 
conversely \cite[p.~64]{Hei27}.'' @  
By the famous $\gamma$ ray microscope 
thought experiment he derived
the 
relation
\beql{Hei27}
\ep(Q)\et(P)\ge\frac{\hbar}{2},
\eeq
where $\ep(Q)$ is the ``mean error'' 
\footnote{Here, ``mean error'' is naturally understood as a notion following that of 
``root-mean-square error'' introduced by Gauss \cite{Gau95}, who introduced and 
called it as the ``mean error'' or the ``mean error to be feared'' in 1821.}
of a position measurement 
and $\et(P)$ is the ``discontinuous change'' (disturbance) thereby caused on the momentum $P$
\cite[p.~64]{Hei27}.
Heisenberg  claimed that \Eq{Hei27} is a ``straightforward mathematical consequence'' of \Eq{CCR}
\cite[p.~65]{Hei27} and
gave its mathematical justification \cite[p.~69]{Hei27},
where Heisenberg 
derived the relation
\beql{Ken27}
\sigma(Q)\sigma(P)\geq\frac{\hbar}{2}
\eeq
for the standard deviations $\si(Q), \si(P)$
\footnote{
The standard deviation is defined for any observable $A$ 
by $\si(A)^2=\bracket{A^2}-\bracket{A}^2$,
where $\bracket{\cdots}$ stands for the mean value in a given state.}
of the position $Q$ and the momentum $P$
\footnote{In Ref. \cite[p.~69]{Hei27} Heisenberg actually derived the relation 
$\tilde{\si}(Q)\tilde{\si}(P)={\hbar}$ 
for $\tilde{\si}(Q)=\sqrt{2}\si(Q)$ and $\tilde{\si}(P)=\sqrt{2}\si(P)$
in Gaussian wave functions.
Kennard \cite{Ken27} derived the relation 
$\tilde{\si}(Q)\tilde{\si}(P)\ge \hbar$ 
that generalizes Heisenberg's relation to arbitrary wave functions.
In the text, we elaborate Heisenberg's original argument
with Kennard's general relation.}.
Heisenberg applied \Eq{Ken27} to the state just after the measurement
assuming the approximate repeatability hypothesis:
(AR) {\em Any measurement with error $\ep(A)$ of an observable $A$ 
leaves the object in the state 
satisfying $\si(A)\le\ep(A)$.}
Under assumption (AR), \Eq{Hei27} 
follows immediately from \Eq{Ken27}
\footnote{ 
It follows immediately from (AR) and \Eq{Ken27} that every simultaneous
measurement of $Q$ and $P$ with error limit $(\ep(Q),\ep(P))$ satisfies 
the relation $\ep(Q)\ep(P)\ge\frac{\hbar}{2}$.
Then, \Eq{Hei27} follows immediately from this relation, since
if $Q$ can be measured with error-disturbance limit $(\ep(Q),\et(P))$,
then $Q$ and $P$ can be simultaneously measured with error limit 
$(\ep(Q),\et(P))$ (see e.g., \cite{03UVR}).}.

Heisenberg's contemporaries supported assumption (AR) 
\footnote{Von Neumann  \cite[pp.~238--239]{vN55} wrote
``We are then to show that if $P, Q$ are two canonically conjugate quantities, 
and a system is in a state in which the value of $P$ can be given with the
accuracy $\ep\{=\si(P)\}$ (i.e., by a $P$ measurement with an error
range $\ep\{=\ep(P)\}$), then $Q$ can be known with no greater accuracy
than $\et\{=\si(Q)\}=h/[4\pi\ep]$. Or: a measurement of $P$ 
with the accuracy $\ep\{=\ep(P)\}$ must bring about an indeterminancy 
$\et\{=\et(Q)\}=h/[4\pi\ep]$ in the value of $Q$.''
Terms in $\{\ldots\}$ are supplemented by the present author.
},
as the ``repeatability hypothesis'' 
\footnote{The repeatability hypothesis is formulated as: 
``If the physical quantity $A$ is measured twice in succession in a system $\bS$, 
then we get the same value each time (Ref.~\cite{vN55}, pp.~335)''
Under this hypothesis, any precise measurement of $A$ changes the state to be 
an eigenstate of the measured observable $A$, which satisfies $\si(A)=0$.
}
was normally assumed for precise measurements.
However, in the light of modern theory of quantum measurement
\cite{DL70,Dav76,84QC,86IQ,89RS,BLM91,BK92,00MN,04URN}, 
the ``repeatability hypothesis'' has been abandoned 
\footnote{Davies and Lewis \cite[p.~239]{DL70} wrote 
``One of the crucial notions is that of repeatability which we show is implicitly
assumed in most of the axiomatic treatments of quantum mechanics, but whose
abandonment leads to a much more flexible approach to measurement theory.''
} and assumption (AR) is no longer accepted.
Thus, 
\Eq{Hei27} 
cannot be considered as an immediate consequence 
of \Eq{Ken27}, although confusions have prevailed 
even in standard text books \cite{vN55,Boh51,Mes59a,Sch68}.

In 1980, Braginsky and coworkers
\cite{BVT80} claimed that
the Heisenberg error-disturbance relation (EDR) \eq{Hei27} leads  to 
a sensitivity limit, called the {\em standard quantum limit} (SQL), for 
gravitational-wave detectors. 
Subsequently, 
Yuen \cite{Yue83} questioned the validity of the SQL
and then
Caves \cite{Cav85} defended the SQL by giving a new proof of the SQL
without directly appealing to \Eq{Hei27}.
Eventually, the conflict was reconciled by pointing out that Caves's derivation of the SQL
still used (unfounded) assumption (AR),
and a solvable model was constructed of an error-free position measurement 
that breaks the SQL \cite{88MS,89RS} (see also Ref.~\cite{Mad88}).
Later,  this model was shown to also break the Heisenberg EDR \eq{Hei27} \cite{02KB5E}
with the uniquely determined notions of rms error and rms disturbance 
for quantum measurements \cite{13DHE}.
Nowadays, the Heisenberg EDR \eq{Hei27} is taken to be a breakable 
limit \cite{BK92,GLM04}, but then the problem remains:
what is the unbreakable constraint between error and disturbance,
which Heisenberg originally intended?  

{\em Universally valid error-disturbance relations.}---In 2003,  
the present author showed the relations
\beqa
\ep(A)\et(B)+\ep(A)\si(B)+\si(A)\et(B) 
&\!\!\ge\!\!&    \frac{1}{2}| \bracket{[A,B]}|,
\label{eq:OEDR}\\
\ep(A) \et(B)+|\bracket{[n(A),B]}+\bracket{[A,d(B)]}|
&\!\!\ge\!\!&  \frac{1}{2}| \bracket{[A,B]}|, \quad
\label{eq:UEDR}
\eeqa
which are universally valid for any observables $A,B$, any system state, and 
any  measuring apparatus,
where standard deviations $\si(\cdots)$ and expectation values $\bracket{\cdots}$
are taken in the state just before measurement, and 
$n(A)$ and $d(B)$ are system observables representing the first moments of 
the error and the disturbance
for $A$ and $B$, respectively
\cite{03HUR,03UVR,03UPQ,04URJ,04URN,05UUP}.
Relation \eq{UEDR} concludes that if the error and the disturbance are statistically 
independent from system state,  then the Heisenberg EDR
\beql{HEDR}
\ep(A)\et(B)\geq\frac{1}{2}|\bracket{[A,B]}|
\eeq
holds, extending the previous results \cite{AG88,Ray94,91QU,Ish91}.
Relation \eq{OEDR} leads to the following new constraints for error-free 
measurements and non-disturbing measurements: if $\ep(A)=0$ then
\beql{DB}
\si(A)\et(B) \ge  \frac{1}{2}\left| \langle  [A,B]  \rangle \right|,
\eeq
and if $\et(B)=0$ then
\beql{EB}
\ep(A)\si(B)\ge  \frac{1}{2}\left| \langle  [A,B]  \rangle \right|,
\eeq
in contrast to that the Heisenberg EDR \eq{HEDR} leads to the divergence of
$\ep(A)$ or $\et(A)$ if $\bracket{[A,B]}\not=0$.
Relation \eq{EB} has been used to derive conservation-law-induced limits for
measurements \cite{03UPQ,04UUP} (see also \cite{02CLU,BL11}),
quantitatively generalizing the Wigner-Araki-Yanase no-go theorem 
\cite{Wig52,AY60,Yan61,91CP} for repeatable measurements under
conservation laws.
Moreover, it has been used to derive an accuracy limit for quantum computing induced 
by conservation laws \cite{03UPQ} 
(see also \cite{02CQC,Ban02b,03QLM,05CQL,06MEP,07CLI,09GFA}).

To derive the above relations,  the ``mean error'' $\ep(A)$ and the disturbance $\et(B)$ 
are defined as follows.  Let us consider a (model of) measuring process 
$(\cK,\kxi,U,M)$ for a system $\bS$ described by a Hilbert space $\cH$
determined by the probe system $\bP$ described 
by a Hilbert space $\cK$,  the initial probe state $\kxi$, 
the unitary evolution $U$ of the composite system $\bS+\bP$ during the measuring interaction,
and the meter observable $M$ of the probe $\bP$  to be directly observed \cite{84QC}.
Then, for observables $A,B$ of $\bS$, the {\em error observable} $N(A)$ 
and the {\em disturbance observable} $D(B)$ 
are defined by
\beqa
N(A)&=&M(\De t)-A(0),\\
D(B)&=&B(\De t)-B(0),
\eeqa
where $A(0)=A\otimes I_{\cK}$, $B(0)=B\otimes I_{\cK}$,
$M(0)=I_{\cH}\otimes M$, $B(\De t)=U^{*}B(0)U$, and 
$M(\De t)=U^{*}M(0)U$.
Then, the {\em (root-mean-square) error} $\ep(A)$ 
and the {\em (root-mean-square) disturbance} $\et(B)$ 
are defined by  \cite{03UVR}
\beqa
\ep(A)^2&\!=\!&\Tr[N(A)^2\rho\otimes\ketbra{\xi}], \\
\et(B)^2&\!=\!&\Tr[D(B)^2\rho\otimes\ketbra{\xi}].
\eeqa

{\em Branciard's relations.}---In 2013, Branciard \cite{Bra13} discussed the tightness of \Eq{OEDR} and
improved \Eq{OEDR} as
\beqa
\lefteqn{\ep(A)^2 \si(B)^2  + \si(A)^2  \et(B)^2 \Bigr.} \notag \\
& & \Bigl. + 2  \ep(A) \et(B) \sqrt{ \si(A)^2\si(B)^2-C_{AB}^2}
\geq C_{AB}^2,
\quad
\label{eq:BEDR}
\eeqa
which is universally valid and stronger than \Eq{OEDR} in the case where $\ep(A)\ne 0$ or
$\et(B)\ne 0$,
where 
\beq
C_{AB}=\frac{1}{2i}\Tr([A,B]\rh).
\eeq
From \Eq{BEDR} he showed that the equality in \Eq{OEDR} cannot be
attained unless $\ep(A)\et(B)=0$.
It was also shown \cite{Bra13} that the above relation can be further strengthened
to be a tight relation for spin measurements in pure states as follows.
Let $A$ and $B$ be 2-valued observables with eigenvalues $\pm1$
and let $\rh$ be a state possibly mixed for which $\bracket{A}=\bracket{B}=0$,
and we further suppose the {\em same spectrum condition} 
that the meter $M$ has the same spectrum as the measured
observable $A$.
In this case, Branciard \cite{Bra13} derived the relation
\beq
\hep(A)^2  + \het (B)^2+ 2  \hep(A) \het(B) \sqrt{ 1-C_{AB}^2}
\geq C_{AB}^2,
\label{eq:BEDRM}
\eeq
where 
\beq
\hep(A)=\ep(A)\sqrt{1-\frac{\ep(A)^2}{4}},~~
\het(B)=\et(B)\sqrt{1-\frac{\et(B)^2}{4}},
\eeq
more stringent than \Eq{BEDR} and showed its tightness for pure input states.

{\em Problem of mixed states.}---In recent papers \cite{14NDQ,Bra14}, 
it is suggested that relation \eq{BEDR} would be considerably strengthened if the measured
system is in a mixed state.
For example, consider the case of spin measurement where $A=Z$, $B=X$ for Pauli operators
$X,Y, Z$ of a spin 1/2 system $\bS$,
the meter $M$ has the same spectrum as $A$, and the input sate is completely random, i.e., $\rh=I_{\cH}/2$.
In this case, \Eq{BEDR} gives no constraint, since $C_{AB}=0$, but a result 
from an information theoretical approach \cite{14NDQ} leads to the relation
\begin{equation} \label{eq:xyent}
\left[\ep(Z)^2+\frac{1}{3} \right]\,\left[\et(X)^2 +\frac{1}{3} \right] \ge\frac{16}{\pi^2 e^2} \approx 0.219. 
\end{equation}
In particular, further consideration concludes 
the relation $\et(B)=\sqrt{2}$ if $\ep(A)=0$.

In what follows, we show a method for strengthening Branciard relations in mixed
states and derive several new universally valid and stronger error-disturbance relations 
in mixed states.  In particular, it is proved that one of them gives an ultimate 
error-disturbance relation for spin measurements, which is tight in any state.

{\em Purification.}---Suppose that the input state $\rh$ has eigenvalues $p_j>0$
with spectral decomposition $\rh=\sum_{j} p_j\ketbra{\ph_j}$.
Then, for any Hilbert space $\cH'$ with $\dim(\cH')\ge \mbox{rank}(\rh)$ describing an 
ancillary system,
we have a ``purification'' $\ket{\Ps}\in\cH\otimes\cH'$ of $\rho$ with the Schmidt
decomposition
$
\ket{\Ps}=\sum_{j}\sqrt{p_j}\ket{\ph_j}\otimes\ket{\et_j},
$
which satisfies
$
\rh=\Tr_{\cH'}[\ketbra{\Ps}],
$
where $\{\ket{\et_j}\}$ is an 
arbitrary orthonormal family in $\cH'$ 
and $\Tr_{\cH'}$ stands for the partial trace over $\cH'$.
In this case, we can extend $A,B$ to $\cH\otimes\cH'$ by
$A'=A\otimes I_{\cH'}$ and $B'=B\otimes I_{\cH'}$,
and extend 
any measuring process $(\cK,\kxi,U,M)$ for $\cH$ to
a measuring process $(\cK,\kxi,U',M)$ for $\cH\otimes\cH'$
by
$
 U'=U\otimes I_{\cH'}.
 $
Then, we easily obtain
\beqa
&\si(A')=\si(A),~~
\si(B')=\si(B),&\\
&\ep(A')=\ep(A),~~ 
\et(B')=\et(B),~~
C_{A'B'}=C_{AB}.\qquad &
\eeqa

Hence, the above extensions preserve relation \eq{BEDR}.
Then, it is easily understood that in the mixed state case \Eq{BEDR} is less stringent
than the pure state case, since the extended measuring apparatus 
$(\cK,\kxi,U',M)$ does not interact with the ancillary system
described by $\cH'$ in the measured system \cite{Bra14}.

{\em Strengthening of error-disturbance relations for mixed 
states.}---In the above discussion, 
the extension conserves all quantities $\si(A)$, $\si(B)$, $\ep(A)$, $\et(B)$, and $C_{AB}$.
Now, we consider another extension using the ``canonical purification'' that
leads to a stronger relation for the quantities $\si(A)$, $\si(B)$, $\ep(A)$, 
and $\et(B)$.

For this purpose a particularly important choice of the ancillary Hilbert space
$\cH'$ is the dual space $\cH^{*}$ of $\cH$,
which consists of all bra vectors $\bra{\ps}\in\cH^{*}$.
Now we suppose $\cH'=\cH^*$.
Then, we define the ``canonical purification'' $\ket{\Ps}\in\cH\otimes\cH^*$ 
of $\rh$ by
\beq
\ket{\Ps}=\sum_{j}\sqrt{p_j}\ket{\ph_j}\otimes\bra{\ph_j},
\eeq
which satisfies
$\rh=\Tr_{\cH^*}[\ketbra{\Ps}]$.
Let 
\beq
{-i}\sqrt{\rh}[A,B]\sqrt{\rh}=W^{\da}|\sqrt{\rh}[A,B]\sqrt{\rh}|
\eeq
be a polar decomposition of the operator $-i\sqrt{\rh}[A,B]\sqrt{\rh}$.
Since $-i\sqrt{\rh}[A,B]\sqrt{\rh}$ is self-adjoint, $W$ can be chosen as
a self-adjoint unitary operator on $\cH$.
Then, the corresponding operator $W^*$ on $\cH^*$
also is a self-adjoint unitary operator.
We extend the observable $B$ on $\cH$ to the observable 
$B'_{W}$ on $\cH\otimes\cH^*$ by
\beq
B'_{W}=(B-\Tr[B\rh]I_{\cH})\otimes W^{*}.
\eeq
Now, we define a new constant  $D_{AB}$ by
\beqa
D_{AB}&=&\frac{1}{2}\Tr(|\sqrt{\rh}[A,B]\sqrt{\rh}|).
\eeqa
Then,  as shown in Supplemental Material  we have 
\beqa
&\si(A')=\si(A),~~ \si(B'_{W})\le \si(B),&\\
&\ep(A')=\ep(A),~~ \et(B_{W}')=\et(B),~~ C_{A'B_{W}'}=D_{AB}.&
\eeqa
Therefore, from \Eq{BEDR} we obtain
\beqa
\lefteqn{
\ep(A)^2 \si(B)^2  + \si(A)^2  \et(B)^2
}\quad\nn\\
& & + 2  \ep(A) \et(B) \sqrt{ \si(A)^2\si(B)^2-D_{AB}^2}
\geq D_{AB}^2.
\label{eq:EDRM}
\eeqa
This relation is equivalent to \Eq{BEDR} if $\rh$ is a pure state,
but significantly stronger than \Eq{BEDR} if $\rh$ is a mixed state;
for $\rh=I_{\cH}/2$, $A=Z$, and $B=X$, we have $C_{AB}=0$ 
but $D_{AB}=1$.

In other words, we consider the physically same measurement $(\cK,\kxi,U',M)$
as the original model $(\cK,\kxi,U,M)$,
but we consider the disturbance on a new observable $B'_{W}$ 
physically different from the original observable $B'=B\otimes I_{\cH'}$.
Then, $B'_{W}$ has the mathematically same value of disturbance $\et(B'_{W})$
as the original value $\et(B)=\et(B')$, and $\si(B'_{W})$ is favorably
smaller than the original value $\si(B)=\si(B')$.
Nevertheless, the lower bound $C_{A'B'}$ can be larger than the original value
$C_{AB}$, which can be maximized up to $D_{AB}$ to obtain the significantly
stringent relation, \Eq{EDRM}, for the original values  $\si(A)$, $\si(B)$, $\ep(A)$, 
and $\et(B)$.

{\em Error-disturbance relation for binary
measurements.}---As the simplest choice of $A$ and $B$, 
the error and disturbance in
spin measurements have been extensively studied in 
theoretically and experimentally \cite{LW10,12EDU,RDMHSS12,
13EVR,13VHE,14ETE,RBBFBW14}.
Let us consider 2-valued observables $A,B$ with eigenvalues $\pm1$
and a state $\rh$ possibly mixed for which $\bracket{A}=\bracket{B}=0$,
and we further suppose the {\em same spectrum condition} 
that the meter $M$ has the same spectrum as the measured
observable $A$.
In this case, Branciard \cite{Bra13} derived  \Eq{BEDRM},
which is more stringent than \Eq{BEDR} and showed its tightness for pure input states.

In this particular case (i.e, $A^2=B^2=I_{\cH}$,  
$M^2=I_{\cK}$, and $\Tr[A\rh]=\Tr[B\rh]=0$), 
as shown in Supplemental Material the canonical purification 
enables us to  strengthen Branciard relation \eq{BEDRM} to the relation
\beqa
\hep(A)^2+ \het (B)^2+ 2  \hep(A) \het(B) \sqrt{1-D_{AB}^2}\geq D_{AB}^2,
\quad
\label{eq:EDRMB}
\eeqa
which is equivalent to \Eq{BEDRM} if $\rh$ is a pure state,
but considerably stronger than \Eq{BEDRM} if $\rh$ is a mixed state.
This strengthening for mixed states will play an important role in applications to 
state estimation problems including quantum cryptographic scenarios.

{\em Tight error-disturbance relation for spin measurements.}---Now we shall show the 
tightness of \Eq{EDRMB} for any state $\rh$ in spin measurements.
Let $\cH\cong\bC^2$ and let $X$, $Y$, and $Z$ be Pauli operators thereon.
We denote by $|0\rangle$ and $|1\rangle$ the eigenstates of $Z$ with 
eigenvalues $+1$ and $-1$, respectively. 
Let $A=Z$ and $B=X$.  
Let us consider a measuring process $(\cK,\kxi,U,M)$
for $\cH$ satisfying the same spactrum condition $M^2=I_{\cK}$.
Suppose that the input state $\rh$ satisfies $\Tr[A\rh]=\Tr[B\rh]=0$.
In this case, we have $\rh=\frac{1}{2}(\al Y+I_{\cH})$ with $-1\le\al\le 1$.
Hence, we have
$
D_{AB}=1.
$

Thus, from \Eq{EDRMB} the relation
\beql{EDRMBS}
(\ep(Z)^2-2)^2+(\et(X)^2-2)^2
\le  4
\eeq
holds for every measuring apparatus 
satisfying the same spectrum condition.

For the case  $\rh=I_{\cH}/2$, we have $C_{AB}=0$ and the Branciard relation,
\Eq{BEDRM}, leads to no constraint. 
It can be easily seen that \Eq{EDRMBS} is significantly stronger than previous
relation \Eq{xyent}
obtained from information theoretical approach \cite{14NDQ}.

In particular, we have the relation
$
\et(X)=\sqrt{2},
$
if $\ep(Z)=0$,  which was also previously obtained in \cite{14NDQ},
so that the precise measurement is attained only with the maximum disturbance,
and conversely we have $\ep(Z)=\sqrt{2}$ if $\et(X)=0$, so that we conclude
{\em no information from no disturbance}.  

Now we shall show that \Eq{EDRMBS} is attained by a measuring process 
uniformly for any state $\rh$.
Suppose that the probe is another spin $1/2$ system
described by the Hilbert space $\cK\cong\bC^2$.
We take the initial probe state as $\kxi=\ket{0'}$ and the meter observable 
in the probe as $M=Z'$; for distinction the prime indicates what defined for $\cK$.
The unitary operator $U$ for the measuring interaction is given by
\beqa\label{eq:U}
U&=&C[X'](I\otimes W(\theta)),
\eeqa
where controlled not $C[X']$ and rotation $W(\theta)$ are given by
\beqa
C[X']&=&\ket{0}\bra{0}\otimes I_{\cK}+\ket{1}\bra{1}\otimes X',\\
W(\theta)&=&\cos\theta Z'+\sin\theta X'.
\eeqa
Then, the rms error $\ep(Z)$ and the rms disturbance $\et(X)$ of the
measuring process $(\bC^2,\ket{0'},U,Z')$ are obtained as 
\beq
\ep(Z)^2=4\sin^2\theta,\quad
\et(X)^2=4\sin^{2}\left( \frac{\pi}{4}-\theta \right)
\eeq
for every input state $\rh$ \cite{13EVR}.
It follows that we have
\beq
(\epsilon(Z)^2-2)^2+(\eta(X)^2-2)^2= 4.
\eeq
Thus, \Eq{EDRMBS} is attained by $(\bC^2,\ket{0'},U,Z')$
with $U$ defined by \Eq{U}
uniformly for any input state $\rh$.

{\em Most general error-tradeoff relation.}---Now we shall consider the most general form of 
the error-disturbance relation.  For this purpose it is convenient to introduce more
general and simpler type of mathematical models. 
A {\em joint measurement model} for a Hilbert space $\cH$ is 
determined by a quadruple $(\cK,\kxi,\X,\Y)$
consisting of a Hilbert space $\cK$,  a unit vector $\kxi\in\cK$, and 
mutually commuting self-adjoint operators $\X,\Y$ on $\cH\otimes\cK$.
For any pair of observables $A,B$ and a density operator $\rh$ on $\cH$, 
the {\em rms errors $\ep(A)$ 
and $\ep(B)$ 
for joint $A,B$ measurement by $(\cK,\kxi,\X,\Y)$ in $\rh$} are defined by
\beqa
\ep(A,\X,\rh)&=&\Tr[(\X-A\otimes I_{\cK})^2\rh\otimes\kb{\xi}]^{1/2},\\
\ep(B,\Y,\rh)&=&\Tr[(\Y-B\otimes I_{\cK})^2\rh\otimes\kb{\xi}]^{1/2}.
\eeqa
Every measuring process $(\cK,\kxi,U,M)$ with $\ep(A)$ and $\et(B)$ is
a joint measurement model $(\cK,\kxi,\X,\Y)$ with $\X=M(\De t)$, $\Y=B(\De t)$,
$\ep(A,\X,\rh)=\ep(A)$, and $\ep(B,\Y,\rh)=\et(B)$.
From now on, we abbreviate $\ep(A)=\ep(A,\X,\rh)$ and $\ep(B)=\ep(B,\Y,\rh)$
when confusions may not occur.

The joint measurement model $(\cK,\kxi,\X,\Y)$ also defines the standard 
deviations 
$\si(\X)$, 
$\si(\Y)$ in the state $\rh\otimes\kb{\xi}$
and the (first moment) biases 
$\de(A)=\Tr[(\X-A\otimes I_{\cK})\rh\otimes\kb{\xi}]$,
$\de(B)=\Tr[(\Y-B\otimes I_{\cK})\rh\otimes\kb{\xi}]$.
Assuming $\si(\X), \si(\Y)\ne 0$, define
\beqas
\lefteqn{E_{\si(\X), \de(A)}(A)
}\quad\nn\\
& \!=\! & 
\sqrt{\si(A)^2-\left( \frac{\si( A)^2 + \si(\X)^2 - (\ep(A)^2 - \de(A)^2)
}{2 \, \, \si(\X)}\right)^{\!2} }, \\
\lefteqn{
E_{\si(\Y), \de(B)}(B) 
}\quad\nn\\
&\! =\! & 
\sqrt{\si(B)^2-\left( \frac{\si( B)^2 
+ \si( {\Y})^2  - (\ep(B)^2 - \de(B)^2)}{2 \,  \, \si(\Y)} \right)^{\!2} }. 
\eeqas
Then, as shown in Supplemental Material we obtain  the relation
\beqa
&  &
E_{\si(\X), \de(A)}(A)^2 \si(B)^2
+ \si(A)^2E_{\si(\Y), \de(B)}(B) ^2 
\nn \\
&+ &
2
E_{\si(\X), \de(A)}(A)  
E_{\si(\Y), \de(B)}(B) 
\sqrt{\si(A)^2\si(B)^2 - D_{AB}^2}\nn\\
&\geq&  D_{AB}^2 .
\quad 
\label{eq:GETRM}
\eeqa
for any given values of $\si(\X)> 0, \si(\Y)> 0$, and $ \de(A),\de(B)$,
which for mixed states strengthens the corresponding relation with $C_{AB}$ 
recently obtained by Branciard \cite{Bra14}.  Branciard \cite{Bra14} showed that his relation 
is stronger than the universally valid error-tradeoff relations previously obtained by
the present author \cite{03UVR}, Hall \cite{Hal04}, and Weston {\em et al.} \cite{WHPWP13}.  
Similarly, we conclude that \Eq{GETRM} is even stronger than those relations enforced by
replacing their lower bound $C_{AB}$ with $D_{AB}$.

{\em Concluding remarks.}---As shown in Supplemental Material 
the canonical purification method can be used to 
strengthen the Robertson relation $\si(A)\si(B)\ge |C_{AB}|$ \cite{Rob29} to obtain
\beql{RH}
\si(A)\si(B)\ge D_{AB}.
\eeq 
Hayashi \cite[p.~194]{Hay06} derived this relation by a different method and  
suggested that $C_{AB}$ in \Eq{OEDR} can be replaced by $D_{AB}$,
but it is not clear whether his method can be used to derive the results obtained 
in this paper.

The rms error $\ep(A)$ is uniquely derived from the classical notion of 
root-mean-square error if $U^{\dag}(I\otimes M)U$ and $A\otimes I$ commute 
as in the case of linear position measurements \cite{13DHE}.
 It is also pointed out in Ref.~\cite{14NDQ} that $\ep(A)$ coincides with the 
root-mean-square error of quantum estimation problems \cite{Hel76} for orthogonal 
families of pure states with the uniform prior distribution,  commonly arising in quantum 
cryptographic protocols.
We adopt the state-dependent approach instead of the state-independent approach
recently advocated by Busch {\em et al.} \cite{BLW13}, since the properties of 
the ``mean error'' can be more suitably described in the state-dependent approach, 
whereas the state-independent approach is more suitable for describing 
the ``worst case error'' \cite{RMHS13}.
In the state-dependent approach, the precise measurement of an observable $A$ 
in a given state $\rh$ is characterized in terms of the rms error as the one which satisfies
$\ep(A)=0$ for all $\ph$ in the cyclic subspace spanned by $A$ and $\rh$ 
\cite{05PCN,06QPC,06NDQ}.  A modification $\overline{\ep}(A)$ of $\ep(A)$ was
proposed in \cite{06NDQ} to satisfy the condition that $\overline{\ep}(A)=0$
if and only if the observable $A$ is precisely measured in the state $\rh$, 
to clear a problem raised by Busch {\em et al.} \cite{BHL04}.
Then, $\overline{\ep}(A)$ satisfies all the relations obtained in this paper, 
by the relation $\overline{\ep}(A)\ge\ep(A)$.
The definition of $\eta(B)$ is derived analogously,  although there are continuing 
debates on alternative approaches \cite{13DHE,WHPWP13,BLW13, RMHS13}.
Further discussions on the significance of the state-dependent approach would be
out of the scope of this paper and will be given elsewhere.

There has been a controversy \cite{Wer04,KS05} on the question about 
experimental accessibility of the error $\ep(A)$ and disturbance $\et(B)$. 
To clear this question two methods have been proposed so far: the
``three-state method''  proposed in Ref.~\cite{04URN} and  the ``weak-measurement method'' 
proposed by Lund-Wiseman~\cite{LW10} based on the relation between 
the rms error/disturbance and the weak joint probability given in Refs.~\cite{88MS,91QU,05PCN}.
Those methods have been experimentally demonstrated in Refs.~\cite{12EDU,13VHE,13EVR,RDMHSS12,14ETE,RBBFBW14}.
The third method using ``two-point quantum correlator'' has been
proposed recently \cite{14DOO}.  
We can expect that those methods will observe the new relations obtained in this paper 
holding even in mixed states as well as their tightness for spin measurements.

\acknowledgments
The author thanks Francesco Buscemi for helpful discussions.
This work was supported by the MIC SCOPE, No.~121806010,
by the John Templeton Foundations, No.~35771,
and by the JSPS KAKENHI, No.~26247016.

\clearpage
\onecolumngrid
 \begin{center}
{\bf\large Supplemental Material}
\end{center}
\bigskip
\twocolumngrid
\setcounter{equation}{0}
\renewcommand{\theequation}{S\arabic{equation}}

\section{Branciard's geometric inequalities}
Let $\cE$ be a real inner product space.
Branciard \cite{Bra13} proved the following three relations
hold for any vectors $\a,\b,\x,\y,\xx,\yy\in\cE$ with $\x\perp\y$, $\xx=\x/\n{\x}$, and $\yy=\y/\n{\y}$.
\begin{widetext}
\beqa
&[\n{\a}^2\!-\!\ip{\a}{\xx}^2]\n{\b}^2\!+\!\n{\a}^2[\n{\b}^2\!-\!\ip{\b}{\yy}^2]
\!+\!2\sqrt{\n{\a}^2\!-\!\ip{\a}{\xx}^2}\sqrt{\n{\b}^2\!-\!\ip{\b}{\yy}^2}\sqrt{\n{\a}^2\n{\b}^2\!-\!\ip{\a}{\b}^2}
\ge \ip{\a}{\b}^2.&\label{eq:BGI1}\\
&\n{\a-\x}^2\n{\b}^2+\n{\a}^2\n{\b-\y}^2
+2\n{\a-\x}\n{\b-\y}\sqrt{\n{\a}^2\n{\b}^2-\ip{\a}{\b}^2}\ge\ip{\a}{\b}^2.&
\label{eq:BGI3}
\eeqa
\end{widetext}

\section{Strengthening of error-disturbance relations for mixed 
states}

In what follows we shall complete the proof of \Eq{EDRM}.

Let $\cH^{*}$ be the dual space of $\cH$.
Then, $\cH^*$ consists of all bra vectors $\bra{\ps}\in\cH^{*}$ 
corresponding to all ket vectors $\ket{\ps}\in\cH$.
The inner product on $\cH^*$ is such that 
$$
(\bra{\xi},\bra{\et})=\bracket{\et|\xi}
$$
for all ket vectors $\kxi,\ket{\et}\in\cH$.

For any operator $A$ on $\cH$, we define the operator $A^*$ on $\cH^*$ by
$$
A^*\bra{\et}=\bra{\et}A
$$
for all $\ket{\et}\in\cH$.  Then,  we have 
$$
(\bra{\xi},A^*\bra{\eta})=\bracket{\eta|A|\xi}
$$
for all $\kxi,\keta\in\cH$.

Now we suppose $\cH'=\cH^*$.
The ``canonical purification'' $\ket{\Ps}\in\cH\otimes\cH^*$ 
of 
$$
\rh=\sum_{j} p_j\ketbra{\ph_j}
$$
is defined by
$$
\ket{\Ps}=\sum_{j}\sqrt{p_j}\ket{\ph_j}\otimes\bra{\ph_j},
$$
which satisfies
$$
\rh=\Tr_{\cH^*}[\ketbra{\Ps}].
$$
The constant  $D_{AB}$ is defined  by
$$
D_{AB}=\frac{1}{2}\Tr(|\sqrt{\rh}[A,B]\sqrt{\rh}|).
$$

Let 
$$
{-i}\sqrt{\rh}[A,B]\sqrt{\rh}=W^{\da}|\sqrt{\rh}[A,B]\sqrt{\rh}|
$$
be a polar decomposition of the operator $-i\sqrt{\rh}[A,B]\sqrt{\rh}$.
Since $-i\sqrt{\rh}[A,B]\sqrt{\rh}$ is self-adjoint, $W$ can be chosen as
a self-adjoint unitary operator on $\cH$.
Then, $W^*$ is a self-adjoint unitary operator on $\cH^*$.
We extend the observable $B$ on $\cH$ to the observable 
$B'_{W}$ on $\cH\otimes\cH^*$ by
$$
B'_{W}=(B-\be I_{\cH})\otimes W^{*},
$$
where $\be=\Tr[B\rh]$.
Then, we have 
$$
\si(B)^2=\bracket{\Ps|{B'}_{W}^2|\Ps}\ge\si(B'_{W})^2.
$$
As in the main text, $A'$ and $U'$ are given by
\beqas
A'&=&A\otimes I_{\cH'},\\
U'&=&U\otimes I_{\cH'}.
\eeqas
Let 
$$
\ket{\Ps'}=\ket{\Ps}\otimes\kxi.
$$
We also have
\begin{widetext}
\beqas
\et(B_{W}')&=&\|B_{W}'(\De t)\ket{\Ps'}-B_{W}'(0)\ket{\Ps'}\|\\
&=&
\| U'^{\da}[(B-\be I_{\cH})\otimes W^*\otimes I_{\cK}]U'\ket{\Ps'}
-(B-\be I_{\cH})\otimes W^*\otimes I_{\cK}\ket{\Ps'} \|\\
&=&
\|U^{\da}[(B-\be I_{\cH})\otimes I_{\cK}]U\otimes W^*\ket{\Ps'}
-(B-\be I_{\cH})\otimes I_{\cK}\otimes W^*\ket{\Ps'}\|\\
&=&
\| [U^{\da}(B\otimes I_{\cK})U-B\otimes I_{\cK}]\otimes W^*\ket{\Ps'}\|\\
&=&
\|I_{\cH}\otimes I_{\cK}\otimes W^*
 [U^{\da}(B\otimes I_{\cK})U -B\otimes I_{\cK}]\otimes I_{\cH^*}\ket{\Ps'}\|\\
&=&
\| [U^{\da}(B\otimes I_{\cK})U-B\otimes I_{\cK}]\otimes I_{\cH^*}\ket{\Ps'}\|\\
&=&
\Tr\{[U^{\da}(B\otimes I_{\cK})U-B\otimes I_{\cK}]^2\rh\otimes \kb{\xi}\}^{1/2}\\
&=&
\et(B).
\eeqas
For $C_{A'B_{W}'}$,  we have
\beqas
2iC_{A'B_{W}'}
&=&\bracket{\Ps|[A',B_{W}']|\Ps}\\
&=&\bracket{\Ps|[A\otimes I_{\cH^*},B\otimes W^*]|\Ps}\\
&=&\bracket{\Ps|[A,B]\otimes W^*]|\Ps}\\
&=&
\Big(\sum_{j}\sqrt{p_j}\ket{\ph_j}\otimes\bra{\ph_j}, 
([A,B]\otimes W^*)\sum_{k}\sqrt{p_k}\ket{\ph_k}\otimes\bra{\ph_k}\Big)\\
&=&
\sum_{j,k}\sqrt{p_j}\sqrt{p_k}\,\Big(\ket{\ph_j}\otimes\bra{\ph_j},
[A,B] \ket{\ph_k}\otimes W^*\bra{\ph_k}\Big)\\
 &=&
\sum_{j,k}\sqrt{p_j}\sqrt{p_k}\,\Big(\ket{\ph_j},[A,B]\ket{\ph_k}\Big)
\Big(\bra{\ph_j},W^*\bra{\ph_k}\Big)\\
&=&
 \sum_{j,k}\sqrt{p_j}\sqrt{p_k}\,\braket{\ph_j|[A,B]|\ph_k}\braket{\ph_k|W|\ph_j}\\
&=&
 \sum_{j}\sqrt{p_j}\,\bra{\ph_j}[A,B]\Big(\sum_{k}\sqrt{p_k}\kb{\ph_k}\Big)W\ket{\ph_j}\\
 &=&
 \sum_{j}\sqrt{p_j}
 \bra{\ph_j}[A,B]\sqrt{\rho}W\ket{\ph_j}\\
&=&
\Tr[W\sqrt{\rh}[A,B]\sqrt{\rho}]\\
&=&
i\Tr[|\sqrt{\rh}[A,B]\sqrt{\rho}|]\\
&=&2i D_{AB}.
\eeqas
\end{widetext}
Thus, in addition to 
\beqas
\si(A')&=&\si(A),\\ 
\ep(A')&=&\ep(A),\\ 
\eeqas
we have
\beqas
\si(B_{W}')&\le&\si(B),\\
\et(B_{W}')&=&\et(B),\\
C_{A'B_{W}'}&=&D_{AB}.
\eeqas
Since  from \Eq{BEDR} we have
\beqas
\lefteqn{\ep(A')^2 \si(B'_{W})^2  + \si(A')^2  \et(B_{W}')^2}\\
& &+ 2  \ep(A') \et(B_{W}') \sqrt{ \si(A')^2\si(B_{W}')^2-
C_{A'B_{W}'}^2}
\geq C_{A'B_{W}'}^2,
\eeqas
we conclude \Eq{EDRM}:
\beqas
\lefteqn{\epsilon (A)^2 \sigma (B)^2  + \sigma (A)^2  \eta (B)^2}\\
& & + 2  \epsilon (A) \eta (B) \sqrt{ \sigma (A)^2\sigma (B)^2-D_{AB}^2}
\geq D_{AB}^2.
\eeqas

\section{Error-tradeoff relation}
In what follows, we shall drive \Eq{ETRM},
the error-tradeoff version of \Eq{EDRM},
from \Eq{BGI3}.

Let $\cH$ be a Hilbert space describing a quantum system $\bS$.
A {\em joint measurement model} for $\cH$ is a quadruple $(\cK,\kxi,\X,\Y)$
consisting of a Hilbert space $\cK$,  a unit vector $\xi\in\cK$, and 
mutually commuting self-adjoint operators $\X,\Y$ on $\cH\otimes\cK$.
For any pair of observables $A,B$ and a density operator $\rh$ on $\cH$, 
the {\em rms errors $\ep(A,\X,\rh)$ and $\ep(B,\Y,\rh)$ 
for joint $A,B$ measurement by $(\cK,\kxi,\X,\Y)$ in $\rh$} are defined by
\beqa
\ep(A,\X,\rh)&=&\Tr[(\X-A\otimes I_{\cK})^2\rh\otimes\kb{\xi}]^{1/2},\\
\ep(B,\Y,\rh)&=&\Tr[(\Y-B\otimes I_{\cK})^2\rh\otimes\kb{\xi}]^{1/2}.
\eeqa

In what follows we shall prove the following theorem.

\begin{Theorem}\label{th:ETRM}
Let $A,B$ be a pair of observables on $\cH$ 
and let $\rh$ be a density operator on $\cH$.
Any joint measurement model $(\cK,\kxi,\X,\Y)$ for $\cH$ satisfy
the relation
\beqa \label{eq:ETRM}
\lefteqn{
\ep(A)^2 \si(B)^2  + \si(A)^2 \ep(B)^2}\quad\nn\\
& & + 2 \ep(A)\ep(B) \sqrt{\si(A)^2\si(B)^2-D_{AB}^2}
\geq D_{AB}^2,
\eeqa
where $\ep(A)=\ep(A,\X,\rh)$ and $\ep(B)=\ep(B,\Y,\rh)$,
and
$\si(A)=\si(A,\rh)$ and $\si(B)=\si(B,\rh)$ 
are the standard deviations of $A,B$ in $\rh$.
\end{Theorem}
\begin{Proof}
Let $\cL(\cW)$ be the space of liner operators on a Hilbert space $\cW$;
if $\cW$ is infinite dimensional, we should consider the space $\cL_{HS}(\cW)$
of  ``Hilbert-Schmidt class'' operators $A$ satisfying  $\Tr[A^*A]<\infty$ 
instead of $\cL(\cW)$,
but for simplicity we abuse the notation $\cL(\cW)$ to denote $\cL_{HS}(\cW)$ 
even in that case.

For any $X,Y\in\cL(\cW)$, define the real number
$\ip{X}{Y}$ by 
\beq
\ip{X}{Y}=\Re\Tr (X^{\da}Y).
\eeq
Then, we have the following:

(i) (Linearity)
$
(Z,xX+yY)=
\Re\Tr Z^{\da}(xX+yY)=x\Re\Tr Z^{\da}X+y\Re\Tr Z^{\da}Y
=x(Z,X)+y(Z,Y).
$

(ii) (Symmetry)
$
(Y,X)=\Re\Tr Y^{\da}X=\Re(\Tr Y^{\da}X)^{*}=\Re\Tr X^{\da}Y=(X,Y).
$

(iii) (Positivity)
$
(X,X)=\Re\Tr[X^{\da}X]\ge0
$

(iv) (Non-degeneracy) If
$
(X,X)=0
$, then
$\Tr[X^{\da}X]=0$, so that $X=0$.
\\
Thus, $\cL(\cW)$ is a real linear space with real-valued inner product 
$\ip{X}{Y}$ and norm $\n{X}=\ip{X}{X}^{1/2}$
for $X,Y\in\cL(\cW)$. 

Let $A,B$ be a pair of observables on $\cH$ 
and let $\rh$ be a density operator on $\cH$.
Let 
$(\cK,\kxi,\X,\Y)$ 
be a joint measurement model for $\cH$.
Let $W$ be a self-adjoint unitary operator in $\cH$ satisfying 
the polar decomposition formula
$$
-iW\sqrt{\rh}[A,B]\sqrt{\rh}=|\sqrt{\rh}[A,B]\sqrt{\rh}|.
$$
Now, we define vectors $\a,\b,\x,\y\in\cL(\cW)$
for $\cW=\cH\otimes\cK$ by
\beqa
 \a&=&A_0\sqrt{\rh}\otimes \kb{\xi},\\
\b&=&-iB_0\sqrt{\rh}W\otimes \kb{\xi},\\
\x&=&\X_0(\sqrt{\rh}\otimes \kb{\xi}),\\
\y&=&-i\Y_0(\sqrt{\rh}W\otimes \kb{\xi}),
\eeqa
where 
\beqa
A_0&=&A-\Tr[A\rh]I_{\cH},\\
 B_0&=&B-\Tr[B\rh]I_{\cH},\\
 \X_0&=&\X- \Tr[A\rh]I_{\cH\otimes\cK},\\
\Y_0&=&\Y-\Tr[B\rh]I_{\cH\otimes\cK}.
\eeqa
Then, we have
\beqas
(\a,\b)&=&\Re\Tr[(A_0\sq)^{\da}(-i)B_0\sq W]\\
&=&\Re(-i\Tr[W\sq A_0 B_0\sq])\\
&=&\Im\Tr[W\sq A_0B_0\sq ]\\
&=&\frac{1}{2i}\Tr( W\sq [A_0,B_0]\sq)\\
&=&\frac{1}{2i}\Tr(W\sq [A,B]\sq)\\
&=&\frac{1}{2}\Tr(|\sqrt{\rh}[A,B]\sqrt{\rh}|).
\eeqas
Thus, we obtain
\beql{DAB}
(\a,\b)=D_{AB}.
\eeq
From
\beqas
(\x,\y)&=&\Re\{-i\Tr[\X_0\Y_0(\sq W\sq \otimes \kb{\xi})]\,\}\\
&=&\Im\Tr[\X_0\Y_0(\sq W\sq \otimes \kb{\xi})]\\
&=&\frac{1}{2i}\Tr\{\,[\X_0,\Y_0](\sq W\sq \otimes \kb{\xi})\,\}\\
&=&0
\eeqas
we have
\beql{MPN}
\x\perp\y.
\eeq

We also have the following relations.
\beqa
\|\a\|^2&=&\Tr[|A_0\sq|^2]=\si(A,\rh)^2,\label{eq:SA}\\
\|\b|^2&=&\Tr[|iB_0\sq W|^2]=\si(B,\rh)^2,\label{eq:SB}\\
\|\x-\a\|^2&=&\Tr[|(\X_0-A_0\otimes I_{\cK})\sq\otimes\kbxi|^2]\nn\\
&=&\ep(A,\X,\rh)^2,\\
\|\y-\b\|^2&=&\Tr[|(\Y_0-B_0\otimes I_{\cK})\sq\otimes \kbxi|^2]\nn\\
&=&\ep(B,\Y,\rh)^2.
\eeqa
Therefore, we obtain \Eq{ETRM}  from \Eq{BGI3}.
\end{Proof}

Let $(\cK,\kxi,U,M)$ be a measuring process for $\cH$
with rms error $\ep(A)$ and rms disturbance $\et(B)$ in a state $\rh$.
Then, $(\cK,\kxi,M(\De t),B(\De t))$ is a joint measurement model
such that 
\beqa
\ep(A,M(\De t),\rh)&=&\ep(A),\\
\ep(B,B(\De t),\rh)&=&\et(B).
\eeqa
Thus, \Eq{EDRM} can also be derived from \Eq{ETRM}.

Note that from the Schwarz inequality 
$$
\|\a\|\, \|\b\|\ge |(\a,\b)|
$$ 
with Eqs.~\eq{DAB}, \eq{SA}, and \eq{SB} we obtain the relation
\beq
\si(A,\rh)\si(B,\rh)\ge D_{AB}
\eeq
for any observable $A,B$ and state $\rh$.  
This strengthen the Robertson inequality \cite{Rob29}
\beq
\si(A,\rh)\si(B,\rh)\ge |C_{AB}|
\eeq
if $\rh$ is a mixed state.

Note also that the tensor product space
$\cW\otimes\cW^*$ is isomorphic to the space $\cL(\cW)$ 
by the correspondence between $\kxi\otimes\bra{\et}\in\cW\otimes\cW^*$ 
and the operator $T_{\kxi\otimes\bra{\et}}=\ket{\xi}\bra{\et}\in\cL(\cW)$ such that
$T_{\kxi\otimes\bra{\et}}\ket{\ps}=\bracket{\et|\ps}\kxi$ for all $\kps\in\cW$.
Moreover, for any density operator $\rh$ on $\cW$, its canonical purification 
$\ket{\Ps}\in\cW\otimes\cW^* $ corresponds to $\sqrt{\rh}\in\cL(\cW)$ under the
above isomorphism.
Thus, for any state $\rh$ the proof using the operator $\sqrt{\rh}\in\cL(\cW)$ 
is essentially transferrable to a proof using the canonical purification 
$\ket{\Ps}\in\cW\otimes\cW^*$ of $\rh$.
We prefer the operator representation $\sqrt{\rh}\in\cL(\cW)$ for 
the canonical purification of $\rh$ because of its mathematical tractability.

\section{Error-disturbance relation for binary measurements}
In this section, we shall derive \Eq{EDRMB} from \Eq{BGI1}.

First, we consider binary joint measurements with the same spectrum condition.

\begin{Theorem}
Let $A,B$ be a pair of observables on $\cH$, 
$\rh$ a density operator on $\cH$,
and $(\cK,\kxi,\X,\Y)$ a joint measurement model for $\cH$.
Suppose that $A^2=B^2=I_{\cH}$,  $\X^2=\Y^2=I_{\cH\otimes\cK}$, and that
$\Tr[A\rh]=\Tr[B\rh]=0$. 
Then, we have the relation
\beql{ETRMB}
\hep(A)^2+ \hep (B)^2+ 2  \hep(A) \hep(B) \sqrt{1-D_{AB}^2}\geq D_{AB}^2,
\eeq
where 
\beqas
\hep(A)&=&\ep(A,\X,\rh)\sqrt{1-\frac{\ep(A,\X,\rh)^2}{4}},\\
\hep(B)&=&\ep(B,\Y,\rh)\sqrt{1-\frac{\ep(B,\Y,\rh)^2}{4}}.\\
\eeqas
\end{Theorem}
\begin{Proof}
Under the assumptions
$A^2=B^2=I_{\cH}$,  $\X^2=\Y^2=I_{\cH\otimes\cK}$, and 
$\Tr[A\rh]=\Tr[B\rh]=0$,
the vectors $\a,\b,\x,\y$ in the proof of Theorem \ref{th:ETRM}
satisfy the relations
\beqa
(\a,\b)&=&D_{AB},\\
\x&\perp&\y,\\
\|\a\|^2&=&\si(A,\rh)^2=1,\\
\|\b|^2&=&\si(B,\rh)^2=1,\\
\|\x\|^2&=&\si(\X,\rh\otimes\kbxi)^2=1,\\
\|\y\|^2&=&\si(\Y,\rh\otimes\kbxi)^2=1,\\
\|\x-\a\|^2&=&\Tr[|(\X_0-A_0)\sq\otimes\kbxi|^2]\nn\\
&=&\ep(A,\X,\rh)^2,\\
\|\y-\b\|^2&=&\Tr[|(\Y_0-B_0)\sq\otimes\kbxi|^2]\nn\\
&=&\ep(B,\Y,\rh)^2.
\eeqa
By the relation $\|\a-\x\|^2=2-2(\a,\xx)$ we have
\beqa
\n{\a}^2\!-\!\ip{\a}{\xx}^2
&=&\|\a-\x\|^2-\frac{\|\a-\x\|^4}{4}\nn\\
&=&\|\a-\x\|^2\left(1-\frac{\|\a-\x\|^2}{4}\right)\nn\\
&=&\ep(A,\X,\rh)^2\left(1-\frac{\ep(A,\X,\rh)^2}{4}\right).
\qquad
\eeqa
Similary, we have
\beqa
\n{\b}^2\!-\!\ip{\b}{\yy}^2
&=&\ep(B,\Y,\rh)^2\left(1-\frac{\ep(B,\Y,\rh)^2}{4}\right).\qquad
\eeqa
Thus, \Eq{ETRMB} follows from \Eq{BGI1}.
\end{Proof}

For the error-disturbance scenario in the text, we conclude \Eq{EDRMB}.
The proof runs as follows.
Let $A,B$ be a pair of observables on $\cH$, 
$\rh$ a density operator on $\cH$,
and $(\cK,\kxi,U,M)$ a measuring process for $\cH$ with $\ep(A)$ and $\et(B)$ in $\rh$.
Then, we have a joint measurement model $(\cK,\kxi,M(\De t),B(\De t))$
such that $\ep(A,M(\De t),\rh)=\ep(A)$ and $\ep(B,B(\De t),\rh)=\et(B)$.
Since $M^2=I_{\cK}$ and $A^2=I_{\cH}$, we have
$M(\De t)^2=B(\De t)^2=I_{\cH\otimes\cK}$.
Thus, \Eq{EDRMB} follows from \Eq{ETRMB}.
 
\section{Error-tradeoff relation with parameters for bias and output fluctuation}
In what follows, we shall derive \Eq{GETRM} from \Eq{BGI1}.

Let $A,B$ be a pair of observables on $\cH$ 
and let $\rh$ be a density operator on $\cH$.
Let 
$(\cK,\kxi,\X,\Y)$ 
be a joint measurement model for $\cH$.
Let $W$ be a self-adjoint unitary operator in $\cH$ satisfying 
the polar decomposition formula
$$
-iW\sqrt{\rh}[A,B]\sqrt{\rh}=|\sqrt{\rh}[A,B]\sqrt{\rh}|.
$$
Now, we define vectors $\a,\b,\x,\y\in\cL(\cW)$
for $\cW=\cH\otimes\cK$ by
\beqa
 \a&=&(A_0\sqrt{\rh})\otimes\kbxi,\\
\b&=&(-iB_0\sqrt{\rh}W)\otimes\kbxi,\\
\x&=&\X_0(\sqrt{\rh}\otimes\kbxi),\\
\y&=&-i\Y_0(\sqrt{\rh}W\otimes\kbxi),
\eeqa
where 
\beqa
 A_0&=&A-\Tr[A\rh]I_{\cH},\\
 B_0&=&B-\Tr[B\rh]I_{\cH},\\
 \X_0&=&\X- \Tr[\X(\rh\otimes\kbxi)]I_{\cH\otimes\cK},\\
\Y_0&=&\Y-\Tr[\Y(\rh\otimes\kbxi)]I_{\cH\otimes\cK}.
\eeqa
Then, by similar calculations to the derivations of \Eq{DAB} 
and \Eq{MPN}, we obtain 
\beqa
(\a,\b)&=&D_{AB},\\
\x&\perp&\y.
\eeqa
We also have the following relations.
\beqa
\|\a\|^2&=&\Tr(\,|A_0\sq|^2\,)=\si(A)^2.\\
\|\b|^2&=&\Tr(\,|-iB_0\sq W|^2\,)=\si(B)^2.\\
\|\x\|^2&=&\Tr[\,|\X_0(\sqrt{\rh}\otimes\kbxi)|^2\,]=\si(\X)^2.\\
\|\y\|^2&=&\Tr[\,|\!-\!i\Y_0(\sqrt{\rh}W\otimes\kbxi)|^2\,]=\si(\Y).^2\qquad
\eeqa
We have
\beqa
\lefteqn{\|\b-\y\|^2}\quad\nn\\
&=&\Tr[\,|\Y_0(\sqrt{\rh}\otimes\kbxi)
-(B_0\sqrt{\rh})\otimes\kbxi|^2\, ]\nn\\
&=&
\Tr[\, |\Y(\sqrt{\rh}\otimes\kbxi)-B\sqrt{\rh}\otimes\kbxi\nn\\
& &-\de(B)(\sqrt{\rh}\otimes\kbxi)|^2\, ]\nn\\
&=&
\ep(B)^2-\de(B)^2.
\eeqa
On the other hand, from
\beqas
\|\b-\y\|^2&=&
\|\b\|^2+\|\y\|^2-2(\b,\y),
\eeqas
we have
\beqa
\lefteqn{
\|\b\|^2-(\b,\yy)^2}\nn\\
&=&
\|\b\|^2-\left(\frac{\|\b\|^2+\|\y\|^2-\|\b-\y\|^2}{2\|\y\|}\right)^2\nn\\
&=&
\si(B)^2-\left(\frac{\si(B)^2+\si(\Y)^2-(\ep(B)^2-\de(B)^2)}{2\si(\Y)}\right)^2 \nn\\
&=&
E_{\si(\Y), \de(B)}(B)^2.
\eeqa
Similarly, we have
\beq
\|\a\|^2-(\a,\xx)^2=E_{\si(\X), \de(A)}(A)^2.
\eeq
Therefore, we obtain \Eq{GETRM}  from \Eq{BGI1}.
\end{document}